\documentclass[journal,draftclsnofoot,onecolumn,12pt]{IEEEtran}

\newif\ifarxiv

\newif\ifonecolumn
\onecolumntrue 

\IEEEoverridecommandlockouts
\usepackage{xurl}
\usepackage{multirow}
\usepackage{dsfont}
\usepackage{subcaption}
\usepackage{cite}
\usepackage{amsmath,amssymb,amsfonts, mathtools}
\usepackage{float}
\usepackage{makecell}
\usepackage{caption}
\usepackage{graphicx}
\usepackage{textcomp}
\usepackage{algorithm}
\usepackage{amsthm}
\usepackage{algpseudocode}
\usepackage{adjustbox}
\usepackage{xcolor}
\usepackage{verbatim}
\usepackage{paralist}
\usepackage{array}
\newcolumntype{?}{!{\vrule width 1.1pt}}
\usepackage{tablefootnote}

\theoremstyle{remark}

\definecolor{green}{rgb}{0.0, 0.5, 0.0} 
\ifonecolumn
\renewcommand{\baselinestretch}{1.5} 
\else
\usepackage[left=0.625in,right=0.625in,bottom=1.1in,top=0.7in]{geometry}
\renewcommand{\baselinestretch}{0.89} 
\fi
\allowdisplaybreaks
\usepackage{acro}
\newcolumntype{?}{!{\vrule width 1pt}}

\DeclareAcronym{snr}{
  short = SNR,
  long = signal-to-noise ratio,}
  \DeclareAcronym{pdf}{
	short = PDF,
	long = probability density function,}
  \DeclareAcronym{sar}{
  short = SAR,
  long = synthetic aperture radar,}

\DeclareAcronym{insar}{
  short = InSAR,
  long = interferometric synthetic aperture radar,}

\DeclareAcronym{ao}{
	short = {AO},
	long = {alternating optimization},
	long-plural-form = {alternating optimizations}
}
\DeclareAcronym{mimo}{
	short = {MIMO},
	long = {multiple-input multiple-output}
}

\DeclareAcronym{uav}{
        short = {UAV},
        long = {unmanned aerial vehicle},
        long-plural-form = {unmanned aerial vehicles}
}

\DeclareAcronym{fdma}{
	short = {FDMA},
	long = {frequency-division multiple-access},
}
\DeclareAcronym{1d}{
	short = {1D},
	long = {one-dimensional},
}
\DeclareAcronym{islr}{
	short = {ISLR},
	long = {integrated sidelobe ratio},
}

\DeclareAcronym{pslr}{
	short = {PSLR},
	long = {peak sidelobe ratio},
}

\DeclareAcronym{3d}{
        short = {3D},
        long = {three-dimensional},
}
\DeclareAcronym{pso}{
	short = {PSO},
	long = {particle swarm optimization},
}

\DeclareAcronym{2d}{
        short = {2D},
        long = {two-dimensional},
}

\DeclareAcronym{dem}{
        short = {DEM},
        long = {digital elevation model},
}
\DeclareAcronym{gs}{
        short = {GS},
        long = {ground station},
        long-plural-form = {ground stations}
}

\DeclareAcronym{los}{
        short = {LOS},
        long = {line-of-sight},
}

\DeclareAcronym{sca}{
        short = {SCA},
        long = {successive convex approximation},
}

\DeclareAcronym{nesz}{
        short = {NESZ},
        long = {noise equivalent sigma zero},
}

\DeclareAcronym{wrt}{
        short = {w.r.t.},
        long = {with respect to},
}
\DeclareAcronym{rhs}{
        short = {r.h.s},
        long = {right-hand side },
}
\DeclareAcronym{gmti}{
	short = {GMTI},
	long = {ground-moving target indication},
}
\DeclareAcronym{lhs}{
        short = {l.h.s},
        long = {left-hand side },
}
\DeclareAcronym{bcd}{
	short = {BCD},
	long = {block coordinate descent},
}
\DeclareAcronym{hoa}{
        short = {HoA},
        long = {height of ambiguity},
}
\DeclareAcronym{cr}{
	short = {CR},
	long = {cognitive radar},
}
\DeclareAcronym{crn}{
	short = {CRN},
	long = {cognitive radar networks},
}
\DeclareAcronym{ml}{
	short = {ML},
	long = {machine learning},
}
\DeclareAcronym{dl}{
	short = {DL},
	long = {deep learning},
}
\DeclareAcronym{ue}{
	short = {UE},
	long = {user equipement},
}
\DeclareAcronym{abs}{
	short = {ABS},
	long = {aerial base station},
}
\DeclareAcronym{ula}{
	short = {ULA},
	long = {uniform linear array},
}
\DeclareAcronym{ati}{
	short = {ATI},
	long = {along-track interferometry},
}
\DeclareAcronym{jsarc}{
	short = {JSARC},
	long = {joint SAR and communication},
}
\DeclareAcronym{roi}{
	short = {RoI},
	long = {region of interest},
}
\DeclareAcronym{awgn}{
	short = {AWGN},
	long = {additive white Gaussian noise},
}
\DeclareAcronym{zf}{
	short = {ZF},
	long = {zero-forcing},
}
\DeclareAcronym{mrt}{
	short = {MRT},
	long = {maximum ratio transmission},
}
\DeclareAcronym{cnr}{
	short = {CNR},
	long = {clutter-to-noise-ratio},
}
\DeclareAcronym{scr}{
	short = {SCR},
	long = {signal-to-clutter-ratio},
}
\DeclareAcronym{an}{
	short = {AN},
	long = {artificial noise},
}
\DeclareAcronym{rcs}{
	short = {RCS},
	long = {radar cross section},
}
\DeclareAcronym{mdp}{
	short = {MDP},
	long = {Markov decision process},
}
\DeclareAcronym{drl}{
	short = {DRL},
	long = {deep reinforcement learning},
}
\DeclareAcronym{td}{
	short = {TD},
	long = {time-division},
}
\DeclareAcronym{jsc}{
	short = {JSAC},
	long = {joint sensing and communication},
}
\DeclareAcronym{csi}{
	short = {CSI},
	long = {channel state information},
}
\DeclareAcronym{isac}{
	short = {ISAC},
	long = {integrated sensing and communication},
}
\DeclareAcronym{ofdm}{
	short = {ODFM},
	long = {orthogonal frequency-division multiplexing},
}
\DeclareAcronym{a2g}{
	short = {A2G},
	long = {air-to-ground},
}
\DeclareAcronym{ppo}{
	short = {PPO},
	long = {proximal policy optimization},
}
\DeclareAcronym{trpo}{
	short = {TRPO},
	long = {trust region policy optimization},
}
\DeclareAcronym{s2c}{
	short = {S2C},
	long = {sensing-to-communication},
}
\begin{document}
	\title{\huge Deep Reinforcement Learning for Cognitive Time-Division Joint SAR and Secure Communications}
		\vspace{-8mm}
	\author{\IEEEauthorblockN{Mohamed-Amine~Lahmeri\IEEEauthorrefmark{1}, Ata Khalili\IEEEauthorrefmark{1}, Yujiao Liu\IEEEauthorrefmark{1}, Anke Schmeink\IEEEauthorrefmark{2}, and
			Robert Schober\IEEEauthorrefmark{1}}\\ 
		\IEEEauthorblockA{\IEEEauthorrefmark{1}Friedrich-Alexander-Universit\"at Erlangen-N\"urnberg (FAU), Germany\\
			\IEEEauthorrefmark{2}RWTH Aachen University Aachen, Germany\\
			}}
	\maketitle 
	\begin{abstract} 
		Synthetic aperture radar (SAR) imaging can be exploited to enhance wireless communication performance through high-precision environmental awareness. However, integrating sensing and communication functionalities in such wideband systems remains challenging, motivating the development of a joint SAR and communication (JSARC) framework. We propose a dynamic time-division JSARC (TD-JSARC) framework for secure aerial communications that is relevant for critical scenarios, such as surveillance or post-disaster communication, where conventional localization of mobile adversaries often fails. In particular, we consider a secure downlink communication scenario where an aerial base station (ABS) serves a ground user (UE) in the presence of a ground-moving eavesdropper. 
		To detect and track the eavesdropper, the ABS uses cognitive SAR along-track interferometry (ATI) to estimate its position and velocity. Based on these estimates, the ABS applies adaptive beamforming and artificial-noise jamming to enhance secrecy. To this end, we jointly optimize the time and power allocation to maximize the worst-case secrecy rate, while satisfying both SAR and communication constraints. Using the estimated eavesdropper trajectory, we formulate the problem as a Markov decision process (MDP) and solve it via deep reinforcement learning (DRL). Simulation results show that the proposed learning-based approach outperforms both learning and non-learning baseline schemes employing equal-aperture and random time allocation. The proposed method also generalizes well to previously unseen eavesdropper motion patterns.
	\end{abstract}
	\section{Introduction}
	Despite mature physical-layer security research \cite{survey}, existing \ac{a2g} methods, such as trajectory optimization, beamforming, and jamming, still face a major challenge in estimating an eavesdropper's location \cite{uav_secure_survey}. Most works either assume perfect channel state information \cite{imperfect_sar1,imperfect_sar2} or employ prior position estimates to model the uncertainty in the eavesdropper's location \cite{imperfect_csi}, which may be difficult to realize in practice. Moreover, conventional radar systems often struggle with strong ground clutter in \ac{a2g} viewing geometries \cite{radar_book}. While \ac{sar} has been identified as a candidate to overcome these limitations \cite{imperfect_sar1, amine1,amine2}, a concrete framework for its practical integration with communication links is not available in the literature.\par 
	Although recent studies have started exploring the potential of \ac{jsarc} systems, the related literature is limited. Existing works have considered \ac{uav}-based bistatic \ac{jsarc} \cite{jsarc1}, energy-efficient trajectory design \cite{jsarc2,jsarc4}, and practical feasibility with \ac{ofdm} signaling \cite{jsarc3}. Nevertheless,  two major gaps persist. First, most studies assume static environments, failing to incorporate cognitive \ac{sar} for dynamic scenarios, such as those involving ground-moving targets \cite{haykin}. Second, the coexistence of \ac{sar} and communication is often oversimplified. For example, the authors of \cite{jsarc2} and \cite{jsarc4} utilize communication signal echoes for \ac{sar}, which leads to degraded imaging performance in practice\cite{sar_tutorial}.\par
	In general, to achieve \ac{jsc}, dual-function systems often rely on complex joint beamforming and waveform design \cite{robust_secure}. Alternatively, \ac{td} \ac{jsc} provides a practical and lower-complexity framework by separating sensing and communication in the time domain. Existing studies have optimized time and power allocation to balance communication and radar detection performance \cite{bi_static_jsc1} and developed scheduling strategies for \ac{td} \ac{jsc} systems\cite{optimal_schedule}. However, these \ac{td} frameworks have been primarily designed for conventional radar and have not yet been extended to \ac{jsarc} systems \cite{radar_book}.\par
	In light of the aforementioned challenges, \ac{ati} emerges as an advanced \ac{sar} technique capable of detecting and tracking ground-moving eavesdroppers even for challenging \ac{a2g} viewing geometries \cite{radar_book}. \ac{ati} exploits the phase difference between signals received by two or more \ac{sar} antennas separated along the flight track \cite{ati_gmti_dual}. This phase difference enables the separation of moving targets from stationary ground clutter, which typically limits the performance of conventional radar systems \cite{radar_book}. \par
	In this work, we propose a cognitive \ac{td} \ac{jsarc} system for secure \ac{a2g} downlink communication. The system leverages \ac{sar} \ac{ati} techniques to estimate the location and velocity of a ground-moving eavesdropper, and uses this information to adapt communication beamforming and \ac{an} jamming. To this end, we formulate a joint time and power allocation problem that maximizes the worst-case average secrecy rate while satisfying both \ac{sar} sensing and communication constraints. The approach is considered {\it cognitive} as parameters such as the \ac{sar} integration time are continuously adapted to the unknown eavesdropper trajectory. The main contributions of this work can be summarized as follows:
	\begin{itemize}
		\item We propose a novel cognitive \ac{td} \ac{jsarc} framework that integrates \ac{sar}-based sensing with secure aerial communications for mitigating ground-moving eavesdroppers.
		\item We develop a dynamic uncertainty model for the eavesdropper location that jointly accounts for \ac{sar} estimation accuracy and prior kinematic constraints, including maximum velocity and acceleration.
		\item We formulate the joint time and power allocation problem as a \ac{mdp} to maximize the average worst-case secrecy rate under both \ac{sar} sensing and communication constraints.
		\item We propose a learning-based solution that requires no prior knowledge of the eavesdropper's trajectory and generalizes effectively to unseen eavesdropper trajectories.
	\end{itemize}
	{\em Notations}: Lower-case letters $x$ denote scalars and boldface lower-case letters $\mathbf{x}$ denote vectors. The set $\{a,\ldots,b\}$ represents the integers between $a$ and $b$. $|\cdot|$ denotes the absolute value of a scalar and the cardinality of a set. Sets $\mathbb{R}^{N \times 1}$ and $\mathbb{C}^{N \times 1}$ represent the $N$-dimensional real and complex vector spaces, respectively. For $\mathbf{x}=(x_1,\ldots,x_N)^\top \in \mathbb{R}^{N \times 1}$ (or $\mathbb{C}^{N \times 1}$), $\|\mathbf{x}\|_2$ denotes the Euclidean norm, $\mathbf{x}^\top$ its transpose, and $\mathbf{x}^\dagger$ its Hermitian. $\max(a,b)$ denotes the maximum of $a$ and $b$, and $[a]^+ \triangleq \max(a,0)$. Furthermore, $\mathcal{CN}(\boldsymbol{\mu},\mathbf{\Sigma})$ denotes a circularly symmetric complex Gaussian distribution with mean $\boldsymbol{\mu}$ and covariance $\mathbf{\Sigma}$, and $\mathbb{E}\{\cdot\}$ denotes the expectation operator. Functions $\operatorname{atan2}(\cdot,\cdot)$, $\sin^{-1}(\cdot)$, and $\tan^{-1}(\cdot)$ denote the two-argument arctangent, inverse sine, and inverse tangent, respectively. Finally, $\mathds{1}\{\cdot\}$ denotes the indicator function.
	\section{System Model} \label{Sec:SystemModel}
\begin{figure}[]
		\centering
		\ifonecolumn
		\includegraphics[width=5in]{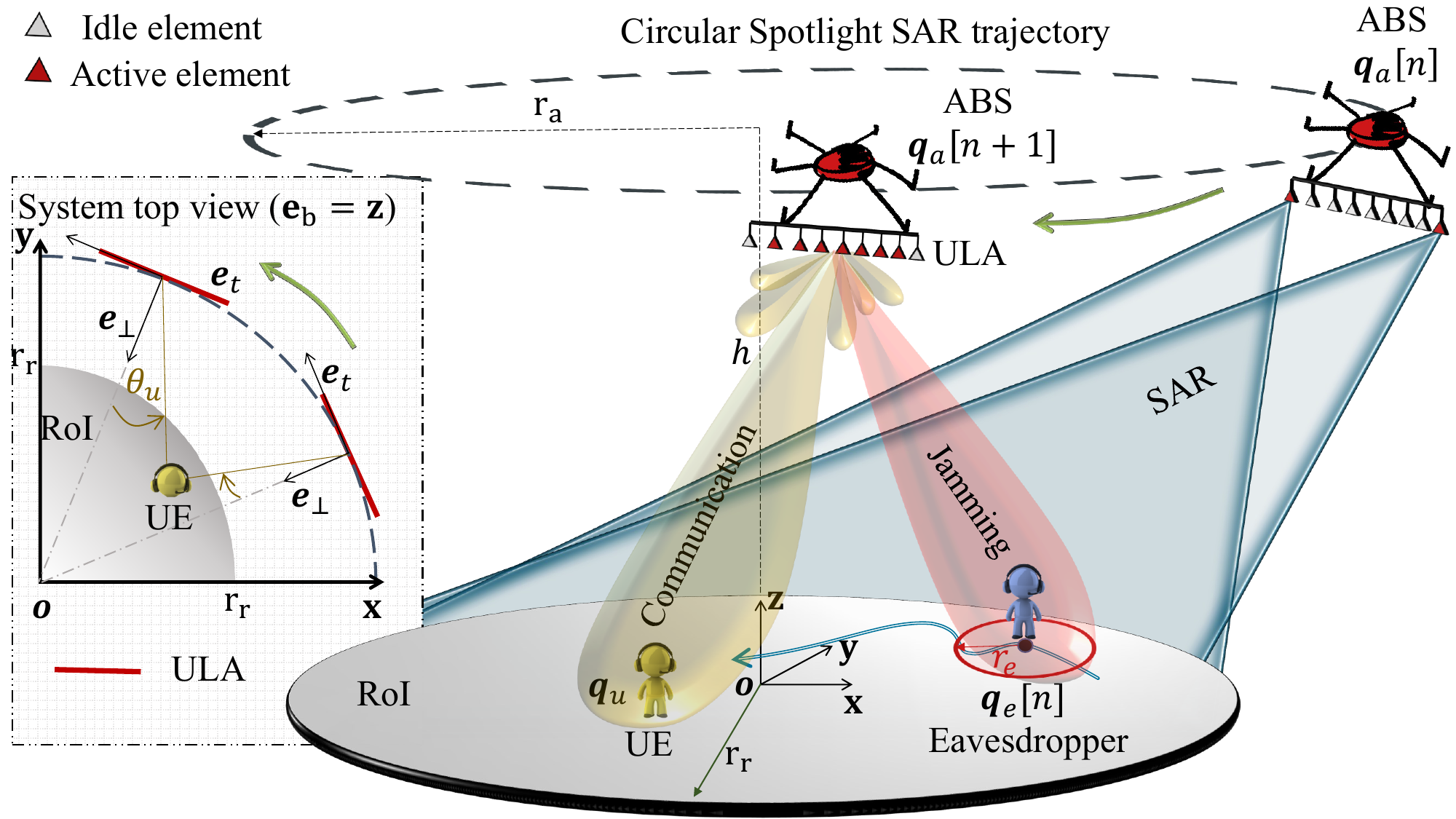}
		\fi
		\caption{System model of an \ac{ula}-equipped \ac{abs} performing \ac{jsarc} in the presence of a ground-moving eavesdropper. The top-view illustrates the Frenet-Serret frame along the \ac{abs} trajectory.}
		\label{fig:system_model}
	\end{figure}

	We consider a downlink scenario where an \ac{abs} serves a single-antenna ground \ac{ue} in the presence of a single-antenna ground-moving eavesdropper. A \ac{3d} Cartesian coordinate system is adopted with origin $\mathbf{o}=(0,0,0)^\top$ at the center of the circular \ac{roi} of radius $r_r$ (see Fig.~\ref{fig:system_model}). The mission duration $T$ is divided into $N>2$ equal slots of duration $\delta_t$, i.e., $T=N\delta_t$, with index set $\mathcal{N}=\{1,\ldots,N\}$. The \ac{abs} is equipped with a \ac{ula} compromising $M_t$ antennas, two of which are dedicated to tracking the eavesdropper via \ac{sar} \ac{ati} \cite{ati_gmti_dual}. The remaining antennas $M_c=M_t-2$ are used for downlink communication with the \ac{ue} located at $\mathbf{q}_u=(q_{u,x},q_{u,y},0)^\top$, as detailed in the subsequent sections.
	\subsection{SAR-based Sensing}
	In each frame, the ABS first performs SAR sensing over a finite integration interval to estimate the eavesdropper’s position and velocity. Based on this estimate, the subsequent communication sub-frame adapts beamforming and jamming to improve secrecy. A longer sensing interval improves localization accuracy, but reduces the time available for data transmission, which creates a sensing-communication tradeoff. Next, we present the proposed dynamic \ac{td} \ac{jsarc} framework.
	\subsubsection{Proposed \ac{td} \ac{jsarc} Framework} 
\begin{figure}[t]
		\centering
		\ifonecolumn
		\includegraphics[width=5in]{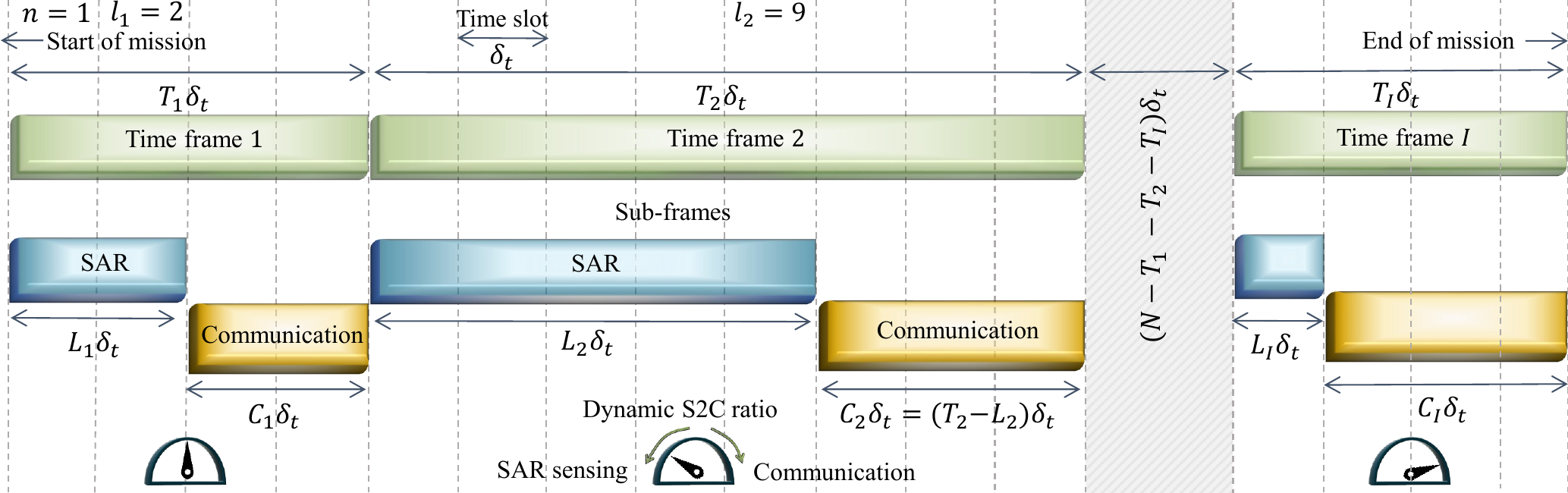}
		\fi
		\caption{Proposed dynamic \ac{td} \ac{jsarc} framework. For illustration, only selected time frames (first, second, and final) are shown, while the gray-colored region represents the omitted intermediate frames.}
		\label{fig:time_allocation}
	\end{figure}

	As shown in Fig.~\ref{fig:time_allocation}, each time frame $i$ with $T_i$ time slots is divided into a sensing sub-frame (i.e., the \ac{sar} aperture\footnote{Consistent with the \ac{sar} literature \cite{sar_tutorial}, the aperture length is expressed in time (i.e., integration time) and can be equivalently converted to its spatial length by scaling with the platform velocity.}) and a communication sub-frame, where $i \in \mathcal{I}=\{1,\dots,I\}$ and $I$ is the total number of apertures.
	Let $\mathcal{T}_i=\{1, \ldots, T_i\}$ be the index set of the $i$-th time frame, where the number of time slots, $T_i$, satisfies:
	\begin{equation}
		2 \le T_i \le N, \quad \forall i \in \mathcal{I}, 
		\quad \text{with} \quad \sum_{i=1}^{I} T_i = N. \label{constraint:time_frame}
	\end{equation}
	The aperture duration of the $i$-th frame is $L_i\delta_t$, where the number of sensing slots $L_i \in \mathbb{N}$ satisfies:
	\begin{equation}
		1 \le L_i \le T_i-1, \quad \forall i \in \mathcal{I}.\label{constraint:aperture_length}
	\end{equation}
	The remaining $C_i=T_i-L_i$ time slots are allocated to communication, and $l_i \in \mathbb{N}$ denotes the last sensing time slot in $\mathcal{T}_i$, see Fig. \ref{fig:time_allocation}. Consequently, the ratio $\frac{L_i}{C_i}$ denotes the \ac{s2c} ratio for the $i$-th sub-frame.
	
	\subsubsection{Proposed Sensing Performance Metrics}        
	During sensing and communication, the \ac{abs} follows a circular trajectory with altitude $h$, velocity $v_a$, and radius $r_a>r_r$ around the \ac{roi}.\footnote{The trajectory is assumed to be pre-optimized to satisfy \ac{sar} requirements such as \ac{snr}, \ac{cnr}, and radar swath \ac{wrt} the \ac{roi}.} The \ac{abs} trajectory is  $\mathbf{q}_a=(\mathbf{q}_a[1],\ldots,\mathbf{q}_a[N])^\top$, where $\mathbf{q}_a[n]=(q_{a,x}[n],q_{a,y}[n],h)^\top$ is the position in time slot $n$ with $q_{a,x}[n]$ and $q_{a,y}[n]$ its horizontal coordinates. During sensing, the \ac{abs} employs circular spotlight \ac{sar} \cite{sar_tutorial}, steering the antenna boresight toward the scene center with the two active antennas separated along the flight direction (see Fig.~\ref{fig:system_model}). Without prior knowledge of the eavesdropper trajectory, the \ac{abs} uses multi-channel \ac{sar} with ground moving target indication and \ac{ati} to separate moving targets from clutter and estimate its position and velocity \cite{gmti_performance}. 
	During the $i$-th sensing sub-frame, a \ac{sar} image with resolution cell approximated by a rectangular area  of size $\delta_r \times \delta_a$ is formed, where the ground-range resolution is $\delta_r = \frac{c}{2 B_r \sin(\eta)}$ \cite{radar_book}. Here, $c$ is the speed of light, $B_r$ the radar bandwidth, and $\eta = \tan^{-1}\!\left(\frac{r_a}{h}\right)$ the incidence angle. The azimuth resolution $\delta_{a}$ depends on the length of the $i$-th aperture, $L_i \delta_t$, and is given 	by\cite{radar_book}:
	\begin{equation}
		\delta_a(L_i) = \frac{\lambda_r r_a}{2 v_a L_i \delta_t}, \forall i \in \mathcal{I},
	\end{equation}
	where $\lambda_r$ is the radar wavelength. A key performance metric for assessing moving-target detectability in clutter-limited environments is the \ac{scr} \cite{gmti_performance, radar_book}:
	\begin{equation}
		\mathrm{SCR}(L_i)\!=\!\frac{\sigma_t}{\sigma_0 \,\delta_r \,\delta_a(L_i)}\!=\!\frac{4 \sigma_t v_a L_i \delta_t  B_r \sin(\eta) }{\sigma_0 c \lambda_r r_a  }, \forall i \in \mathcal{I},
	\end{equation}
	where $\sigma_t$ denotes the radar cross section of the target (i.e., the eavesdropper) and $\sigma_0$ is the normalized terrain backscatter coefficient. Note that the \ac{scr} is independent of the radar transmit power which is assumed to be fixed as the \ac{abs} trajectory is pre-optimized. 
	\subsubsection{Eavesdropper Uncertainty Model}
\begin{figure}[]
		\centering
		\ifonecolumn
		\includegraphics[width=4in]{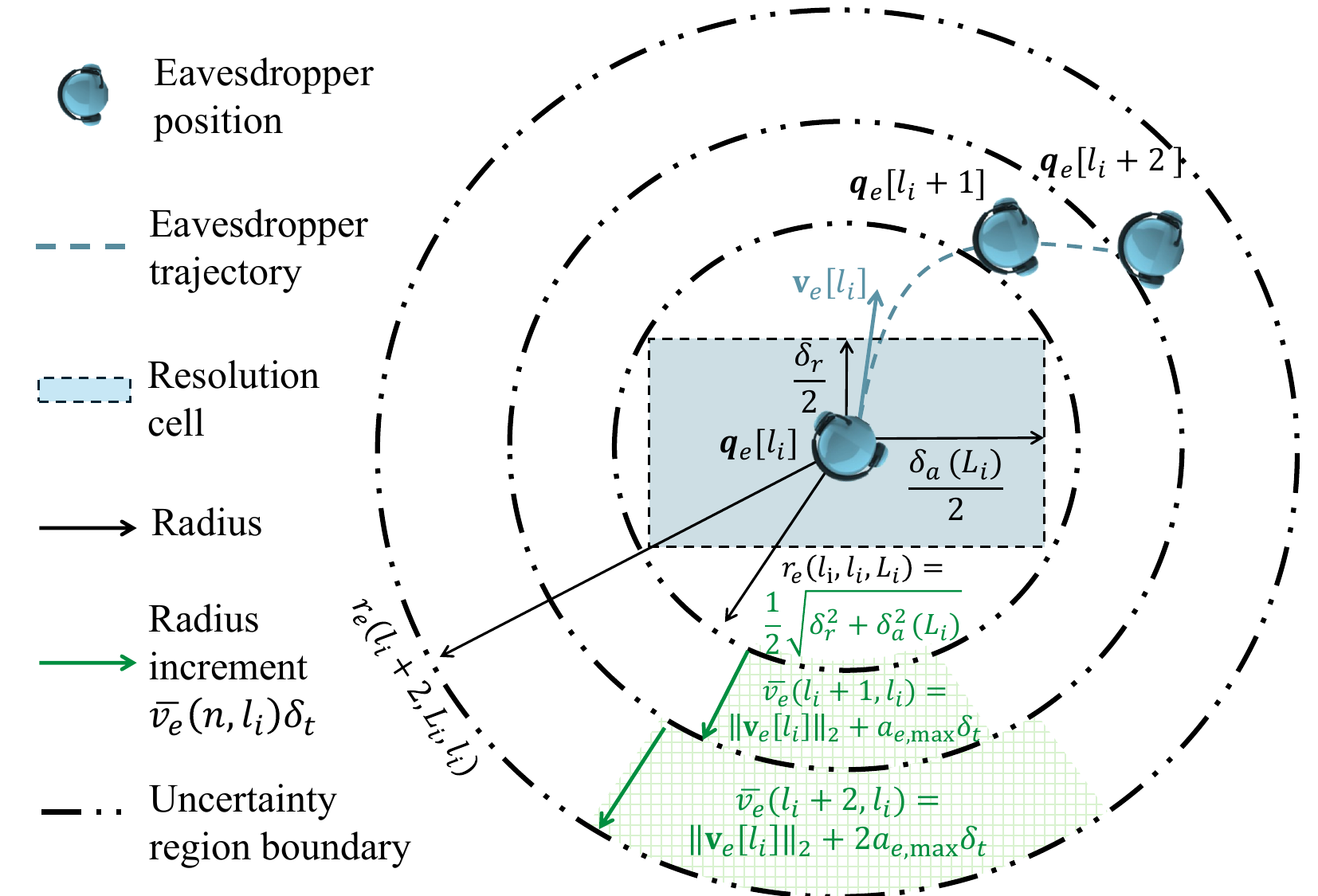}
		\fi
		\caption{Top-view illustration of the evolution of the eavesdropper uncertainty region with radius $r_e$ and a simplistic geometric representation of the \ac{sar} resolution cell.}
		\label{fig:uncertainty_model}
	\end{figure}

	The eavesdropper follows the trajectory $\mathbf{q}_e$ $= (\mathbf{q}_e[1],$ $\ldots, \mathbf{q}_e[N])^\top$, where $\mathbf{q}_e[n]$ denotes its instantaneous \ac{3d} position. The corresponding velocity and acceleration are denoted by  $\mathbf{v}_e[n]=\frac{\mathbf{q}_e[n+1]-\mathbf{q}_e[n]}{\delta_t}, \forall n \in \mathcal{N}\setminus \{N\}$ and  $\mathbf{a}_e[n]=\frac{\mathbf{v}_e[n+1]-\mathbf{v}_e[n]}{\delta_t}, \forall n \in \mathcal{N}\setminus \{N-1, N\}$, respectively. In the first communication time slot of the $i$-th time frame, i.e., $l_i+1$,  both the trajectory and velocity of the eavesdropper are \emph{a priori} unknown. The \ac{abs} only knows (i) the maximum eavesdropper speed $v_{e,\mathrm{max}}$, (ii) the maximum acceleration $a_{e,\mathrm{max}}$, (iii) the previously estimated position $\mathbf{q}_e[l_i]$, and (iv) the previously estimated eavesdropper velocity $\mathbf{v}_e[l_i]$.\footnote{Note that (i) and (ii) are target dependent (e.g., car or person), whereas the estimates (iii) and (iv) are obtained from the previous sensing sub-frame.} 
	Based on this information, the \ac{abs} defines an uncertainty region modeled as a circular disk centered at $\mathbf{q}_e[l_i]$ (see Fig. \ref{fig:uncertainty_model}). The radius of this region evolves over time, and is given for each slot $n \in \mathcal{T}_i$ by:
	\begin{equation}
		r_e(n,l_i,L_i)=\frac{\sqrt{ \delta^2_r+ \delta^2_a(L_i-[l_i-n]^+)}}{2}+[n-l_i]^+ \overline{v_e}(n,l_i)\delta_t, \label{eq:uncertainty_radius}
	\end{equation}
	where $\overline{v_e}(n,l_i)$ denotes an upper bound on the eavesdropper speed in time slot $n < l_i$, derived from the previous velocity estimate and acceleration limits as follows:
	\begin{align}
		&\overline{v_e}(n,l_i) =\max\left(	\|\mathbf{v}_e[l_i]\|_2 + (n-l_i) a_{e,\mathrm{max}},v_{e,\mathrm{max}} \right).\label{eq:upper_bound_velocity}
	\end{align}  
 Note that, based on (\ref{eq:uncertainty_radius}) and (\ref{eq:upper_bound_velocity}), in time slot $l_i$, the eavesdropper location is known up to the \ac{sar} resolution cell of the $i$-th aperture, i.e., $r_e(l_i,l_i,L_i)=\frac{1}{2}\sqrt{\delta_r^2+\delta^2_a(L_i)}$, as illustrated by the smallest disk in  Fig. \ref{fig:uncertainty_model}. Consequently, the uncertainty region estimated by the \ac{abs} in time slot $ n \in \mathcal{T}_i, \forall i \in \mathcal{I},$ is given by: 
\begin{align}
	\adjustbox{scale=0.94}{$
		\mathcal{R}^n_i = \Bigl\{ \mathbf{q} \in \mathbb{R}^{3 \times 1} \,\Big|\, 
		\|\mathbf{q} - \mathbf{q}_e[l_i]\|_2 \le r_e(n,l_i,L_i),\,
		z(\mathbf{q}) = 0 \Bigr\}
		$}
\end{align}
	where $z(\mathbf{q})$ denotes the altitude of $\mathbf{q}$. Since $||\mathbf{v}_e[n]||_2\leq \overline{v_e}(n,l_i),  \forall l_i < n \in \mathcal{T}_i$, the worst-case eavesdropper location lies on the boundary of $\mathcal{R}_i^n$.  Note that high eavesdropper speeds or short \ac{sar} integration times degrade localization accuracy, while longer integration improves resolution (smaller $\mathcal{R}_i^n$) but reduces communication time.
	\subsection{Downlink Communication}
	After each sensing aperture, the \ac{abs} establishes downlink communication with the \ac{ue} in the presence of the eavesdropper. In this sub-section, all communication-related variables are provided for time slot $n \in \mathcal{T}_i$, where $n > l_i$, $\forall i \in \mathcal{I}$.
	\subsubsection{Transmitted Signal}
	The signal transmitted by the \ac{abs} in time slot $n$ is given by: 
	\begin{equation}
		\mathbf{x}[n]=\mathbf{w}[n]u[n]+\mathbf{a}[n]. 
	\end{equation}
	Here, $\mathbf{w}[n]\in \mathbb{C}^{M_c \times 1}$ denotes the \ac{abs} transmit beamforming vector, $u[n] \in \mathbb{C}$ is the information symbol with $\mathbb{E}\{|u[n]|^2\}=1, \forall n$, and $\mathbf{a}[n]\in \mathbb{C}^{M_c \times 1} \sim \mathcal{C}\mathcal{N}(\mathbf{0}, \mathbf{A}[n])$ represents the \ac{an} with covariance $\mathbf{A}[n]\in \mathbb{C}^{M_c\times M_c}$.
	\subsubsection{Received Signal}
	To model antenna orientation along the trajectory, a Frenet– Serret frame $(\mathbf{e}_\bot[n],\mathbf{e}_t[n],\mathbf{e}_b)$ is attached in each time slot $n$ for the \ac{abs}, such that $\mathbf{e}_\bot[n]=-\frac{1}{r_a}\mathbf{q}_a[n]$, $\mathbf{e}_t[n]=\frac{1}{r_a}(-q_{a,y}[n],\,q_{a,x}[n],\,0)^\top$, and $\mathbf{e}_b=(0,0,1)^\top$ (cf. Fig.~\ref{fig:system_model}).
	In this local coordinate system, the array steering vector $\mathbf{v}\in\mathbb{C}^{M_c\times 1}$ is given by:
	\begin{equation}
		\adjustbox{scale=0.98}{$\mathbf{v}(\theta)\!=\!\left(1,\!e^{-j\pi\sin(\theta)},\!e^{-j2\pi\sin(\theta)},\!\ldots,\!e^{-j(M_t-3)\pi\sin(\theta)}\!\right)^\top\!,$}
	\end{equation}
	where $\theta$ is the angle between the signal direction and the array broadside aligned with $\mathbf{e}_\bot$. Assuming a \ac{los} dominated scenario and a free-space path-loss model, the \ac{abs}-to-\ac{ue} and \ac{abs}-to-eavesdropper channels in time slot $n$ are respectively given by:
	\begin{align}
		\mathbf{h}_u[n] &= \frac{\sqrt{\beta_0}\, \mathbf{v}(\theta_u[n])}{d_u[n]}, \quad d_u[n]=\|\mathbf{q}_a[n] - \mathbf{q}_u\|_2, \label{eq:user_channel}\\
		\mathbf{h}_e[n] &= \frac{\sqrt{\beta_0}\, \mathbf{v}(\theta_e[n])}{d_e[n]}, \quad d_e[n]= \|\mathbf{q}_a[n] - \mathbf{q}_e[n]\|_2,\label{eq:eavesdropper_channel}
	\end{align}
	where $\beta_0$ denotes the channel power gain at a reference distance of 1 m. Angles $\theta_u$ and $\theta_e$ are given by: 
	\begin{align}
		&\theta_u[n]\!=\!\mathrm{atan2}  \Big(\! (\mathbf{q}_u\!-\!\mathbf{q}_a[n])^\top\!\mathbf{e}_t[n], (\mathbf{q}_u\!-\!\mathbf{q}_a[n])^\top\!\mathbf{e}_\bot[n]\Big)\!, \label{eq:user_azimuth} \\
		&\adjustbox{scale=0.93}{$\theta_e[n]\!=\!\mathrm{atan2} \Big(\! (\mathbf{q}_e[n]\!-\!\mathbf{q}_a[n])^\top\!\mathbf{e}_t[n], (\mathbf{q}_e[n]\!-\!\mathbf{q}_a[n])^\top\!\mathbf{e}_\bot[n]\!\Big)$}\label{eq:eavesdropper_azimuth}\!.
	\end{align} 
	The received signals at the \ac{ue} and the eavesdropper in time slot $n$ are given by:
	\begin{align}
		y_u[n]&= \mathbf{h}^\dagger_u[n] \mathbf{w}[n] u[n] +  \mathbf{h}^\dagger_u[n]\mathbf{a}[n] +n_u[n],  \\
		y_e[n]&= \mathbf{h}^\dagger_e[n] \mathbf{w}[n] u[n] +\mathbf{h}^\dagger_e[n]\mathbf{a}[n]+ n_e[n],
	\end{align}
	where $n_u[n] \sim \mathcal{CN}(0,\sigma_u^2)$ and $n_e[n] \sim \mathcal{CN}(0,\sigma_e^2)$ denote the additive white Gaussian noise at the \ac{ue} and eavesdropper with variances $\sigma_u^2$ and $\sigma_e^2$, respectively.
	\subsubsection{Beamforming and Jamming}
	The \ac{abs} jams the last estimated eavesdropper position (i.e., the center of $\mathcal{R}_i^n$), $\mathbf{q}_e[l_i]$,  by directing \ac{an} toward this location:
	\begin{equation}
		\mathbf{A}[n]= \alpha[n] P_{\mathrm{com}}^{\mathrm{max}} \frac{\mathbf{h}_e[l_i] \mathbf{h}_e^\dagger[l_i]}{\|\mathbf{h}_e[l_i]\|_2^2},
	\end{equation}
	where $\alpha[n] \in [0, 1]$ is a power-splitting factor and $P_{\mathrm{com}}^{\mathrm{max}}$ is the maximum communication power. For downlink transmission, \ac{mrt} beamforming is adopted to maximize the received \ac{snr} at the user, yielding:
	\begin{equation}
		\mathbf{w}[n]= \sqrt{(1-\alpha[n])P_{\mathrm{com}}^{\mathrm{max}}}\frac{\mathbf{h}_u[n]}{\|\mathbf{h}_u[n]\|_2}.
	\end{equation}
	Note that, due to the unknown instantaneous eavesdropper position, the AN is steered toward the latest available estimate $\mathbf{q}_e[l_i]$. 
	\subsubsection{Secrecy Rate}
	The achievable rate of the \ac{ue} (bits/s/Hz) in time slot $n$ is expressed as follows: 
	\begin{align}
		R_u[n] = 	\log_2 \Biggl( 1 + \frac{\Big|\mathbf{w}^\dagger[n]\mathbf{h}_u[n]\Big|^2}{\mathbf{h}^\dagger_u[n]\mathbf{A}[n]\mathbf{h}_u[n]+\sigma_u^2} \Biggr),
	\end{align}
	Similarly, the achievable rate of the eavesdropper is given by:
	\begin{align}
		R_e[n] =	\log_2 \Biggl( 1+\frac{\Big|\mathbf{w}^\dagger[n]\mathbf{h}_e[n]\Big|^2}{\mathbf{h}^\dagger_e[n]\mathbf{A}[n]\mathbf{h}_e[n]+\sigma_e^2}\Biggr),
	\end{align}
	Consequently, the worst-case secrecy rate between the \ac{abs} and the \ac{ue} is given by:
	\begin{equation}
		R[n]\!=\!\left[R_u[n]\!-\!\max_{\mathbf{q}_e[n] \in \mathcal{R}^n_i } R_e[n](\mathbf{q}_e[n])\!\right]^+\!, \forall l_i < n \in \mathcal{T}_i. \label{eq:rate}
	\end{equation}
	According to the proposed \ac{td} \ac{jsarc} framework, for sensing slots $n \le l_i$, we have $R[n]=R_u[n]=R_e[n]=0$.
	\section{Problem Formulation and Solution}
	We aim to maximize the average worst-case secrecy rate by jointly optimizing the time and power allocation. In particular, the scheduling variables include the total number of frames (i.e., the \ac{sar} apertures) $I$, the length of each frame $\{T_i\}_{i=1}^I$, and the length of each sensing aperture $\{L_i\}_{i=1}^I$. The resulting optimization problem is formulated as follows: 
	\begin{subequations}
		\begin{alignat}{2}
			&(\mathcal{P}): \quad && \mathop{\text{maximize}}_{\substack{I,\, T_i,\, L_i \in \mathbb{N}, \forall i \in \mathcal{I}\\ \alpha[n] \in [0, 1], \forall n \in \mathcal{N}}} 
			\quad \frac{1}{N}\sum_{n=1}^{N} R[n] \notag \\[2mm]
			&  && \text{subject to } (\ref{constraint:time_frame}),\ (\ref{constraint:aperture_length}), \notag \\ 
			& && \mathrm{SCR}(L_i) \ge \mathrm{SCR}_{\min}, \quad \forall i\in\mathcal{I}, \label{constraint:scr}\\
			& && \frac{1}{N}\sum_{n=1}^{N} R_u[n] \ge R_{\min}. \label{constraint:minimum_user_rate}
		\end{alignat}
	\end{subequations}
	Constraint (\ref{constraint:scr}) ensures a minimum \ac{scr} to separate target motion from strong clutter, while constraint (\ref{constraint:minimum_user_rate}) guarantees a minimum average rate for the legitimate user. 
	Problem $(\mathcal{P})$ captures a key trade-off in \ac{td} \ac{jsarc}: Longer \ac{sar} apertures improve spatial resolution and sensing accuracy, enhancing communication in later frames, but reduce transmission time. Moreover, the power-splitting factor $\alpha$ balances information transmission via \ac{mrt} beamforming and \ac{an} jamming.\par
	Problem $(\mathcal{P})$ is a challenging non-convex mixed-integer optimization problem. First, the eavesdropper trajectory is unknown to the \ac{abs}. Second, the number of \ac{sar} apertures $I$, which determines the problem dimension, is itself an optimization variable. Last, the problem involves both continuous and integer variables, whose strong coupling makes global optimization computationally prohibitive.\par
	To address problem $\mathcal{(P)}$, we reformulate it as an \ac{mdp} and solve it using \ac{drl}. This approach is motivated by the unknown trajectory of the eavesdropper. Additionally, \ac{drl} has shown great potential for solving complex and high-dimensional decision-making problems \cite{rl_book}.
	\subsection{Markov Decision Process Framework}
	An \ac{mdp} is defined by the tuple
	$\mathcal{M} = (\mathcal{S}, \mathcal{A}, r, \gamma)$,
	where $\mathcal{S}$ denotes the state space, $\mathcal{A}$ the action space,
	$r$ the reward function, and $\gamma \in [0,1)$ the discount factor \cite{rl_book}. 
	\subsubsection{Action}
	The original time-allocation variables $I$, $\{T_i\}_{i=1}^I$, and $\{L_i\}_{i=1}^I$ can be equivalently represented via a binary scheduling action $a[n] \in \{0,1\}$. Specifically, at time slot $n$, the agent selects sensing for $a[n]=1$ or communication for $a[n]=0$. Given the full action sequence  $\{a[n]\}_{n=1}^{N}$, the corresponding time-allocation variables $I$, $\{T_i\}_{i=1}^I$, and $\{L_i\}_{i=1}^I$ can be reconstructed accordingly. To maintain a discrete action space,  the power-splitting factor $\alpha[n]$ (relevant only for $a[n]=0$) is excluded from the agent's action and is addressed separately via a low-dimensional robust optimization problem. To this end, the instantaneous worst-case secrecy rate, $R[n]$, is maximized for a given time-allocation, leading to the following max-min problem:				
	\begin{alignat}{2}
		&(\mathcal{P}_1): \quad && \mathop{\text{max}}_{\alpha[n] \in [0, 1]}  \mathop{\text{min}}_{\mathbf{q}_e[n] \in \mathcal{R}^n_i} 
		\quad  \Big[ R_u[n]- R_e[n]\Big]^+. \notag
	\end{alignat}
	We solve the low-dimensional problem, $(\mathcal{P}_1)$, via finite exhaustive search over $\alpha[n]\in[0,1]$. For a fixed $\alpha[n]$, the inner minimization is equivalent to maximizing the eavesdropper communication \ac{snr}, which can be expressed in terms of $d_e[n]$ and $\theta_e[n]$ (see \eqref{eq:eavesdropper_channel} and \eqref{eq:eavesdropper_azimuth}, respectively) as follows:
	\begin{alignat}{2}
		&(\mathcal{P}_2): \quad &&  \mathop{\text{max}}_{\theta_e[n], d_e[n]} 
		\quad \frac{|\mathbf{h}^\dagger_u[n]\mathbf{v}(\theta_e[n])|^2} {|\mathbf{h}^\dagger_e[l_i]\mathbf{v}(\theta_e[n])|^2+d^2_e[n] \mathcal{X}_n} \label{prob:p2}\\[2mm]
		&  && \text{subject to }  \notag (\ref{eq:eavesdropper_channel}), (\ref{eq:eavesdropper_azimuth}), \notag
	\end{alignat}
	where $\mathcal{X}_n\triangleq\frac{\sigma_e^2 d_e^2[n]||\mathbf{h}_e[l_i]||_2^2}{\beta_0 \alpha[n] P_{\mathrm{com}}^{\mathrm{max}}}$. To simplify notation, let $\hat{d}_{e,n} \triangleq ||\mathbf{q}_a[n]-\mathbf{q}_e[l_i]||_2$ be the distance between  the \ac{abs} in time slot $n$ and the most recently estimated eavesdropper location. Similarly, $\hat{\theta}_{e,n}$  is obtained by replacing $\mathbf{q}_e[n]$ with $\mathbf{q}_e[l_i]$ in (\ref{eq:eavesdropper_azimuth}). Since $\mathbf{q}_e[n] \in \mathcal{R}^n_i, \forall n$, then, the worst-case azimuth angles belong to set $[\hat{\theta}_{e,n}-\Delta\theta_e[n], \hat{\theta}_{e,n}+\Delta\theta_e[n]]$:
	\begin{align}
		\Delta\theta_e[n]\!=\!
		\begin{cases}
			\!\pi\!, &\text{if}\; \hat{d}_{e,n}\le r_e(n,l_i,L_i),\\
			\sin^{-1}\!\left(\frac{r_e(n,l_i, L_i)}{\hat{d}_{e,n}}\right), &\text{otherwise}.
		\end{cases}	\label{eq:theta_uncertainty}
	\end{align}
	For a given $\theta_e[n]$, $d_e[n]$, the worst-case distance corresponds to the minimum distance between the \ac{abs} and the uncertainty region, since the objective function in \eqref{prob:p2} decreases with increasing $d_e[n]$. Using the law of cosines, the distance $d_e[n]$ is obtained as the smaller non-negative root of: 
	\begin{equation}
		d^2_e[n] - 2\hat{d}_{e,n} \cos\!\big(\theta_e[n]-\hat{\theta}_{e,n}\big)\, d_e[n] + \hat{d}_{e,n}^{\,2}- r^2_e(n,L_i,l_i)= 0.
		\label{eq:de_quadratic}
	\end{equation}
	This reduces the inner minimization in $(\mathcal{P}_2)$ to a one-dimensional search over $\theta_e[n]$.
	To summarize, we derive an $\epsilon$-optimal solution to problem $(\mathcal{P}2)$  based on a \ac{2d} exhaustive grid search over $\alpha[n]$ and $\theta_e[n]$ with respective precisions $\epsilon_{\alpha}$ and $\epsilon_{\theta}$. This results in a polynomial-time complexity of $\mathcal{O}\!\left(\frac{1}{\epsilon_{\alpha}\epsilon_{\theta}}\right)$ per agent step, where $\mathcal{O}(\cdot)$ is the big-O notation. The resulting procedure for computing the robust power-splitting factor $\alpha^\star[n]$ is summarized in \textbf{Algorithm}~\ref{alg:alpha_search}.
	\ifonecolumn
	\begin{algorithm}[t]
		\caption{Robust Power Allocation}\label{alg:alpha_search}
		\begin{algorithmic}[1]
			\State \textbf{Input:} Slot index $n$ with $a[n]=0$ and given $l_i$, uncertainty radius $r_e(n,L_i,l_i)$, last eavesdropper estimates $(\hat{d}_{e,n},\hat{\theta}_{e,n})$ at $l_i$, and grid step $\epsilon$.
			\State \textbf{Output:} $\epsilon$-optimal power-splitting factor $\alpha^\star[n]$ as  solution to $(\mathcal{P}_2)$.
			\State Set $\mathrm{best}=0$ and compute $\Delta\theta_e[n]$ according to (\ref{eq:theta_uncertainty})
			\State \textbf{for} $\alpha\in \{0,\epsilon,2\epsilon,\ldots,1\}$ \textbf{do}
			\State\hspace{\algorithmicindent} Set $\mathrm{worst}=\infty$.
			\State\hspace{\algorithmicindent} \textbf{for} $\theta_e\in\{\hat{\theta}_{e,n}-\Delta\theta_e[n],\hat{\theta}_{e,n}-\Delta\theta_e[n]+\epsilon,\ldots,\hat{\theta}_{e,n}+$ \State\hspace{\algorithmicindent} $\Delta\theta_e[n]\}$ \textbf{do}
			\State\hspace{\algorithmicindent}\hspace{\algorithmicindent} Compute $d_e$ by solving (\ref{eq:de_quadratic})
			\State\hspace{\algorithmicindent}\hspace{\algorithmicindent} Construct candidate $\mathbf{q}^{\mathrm{worst}}_e$ from $d_e$ and $\theta_e$
			\State\hspace{\algorithmicindent}\hspace{\algorithmicindent} Compute $R[n]$ for $\mathbf{q}_e[n]= \mathbf{q}^{\mathrm{worst}}_e$ and $\alpha[n]=\alpha$.
			\State\hspace{\algorithmicindent}\hspace{\algorithmicindent} \textbf{if} $R[n]<\mathrm{worst}$ \textbf{then} set $\mathrm{worst}=R[n]$
			\State\hspace{\algorithmicindent} \textbf{end for}
			\State\hspace{\algorithmicindent} \textbf{if} $\mathrm{worst}>\mathrm{best}$ \textbf{then} set $\mathrm{best}=\mathrm{worst}$ and $\alpha^\star[n]=\alpha$.
			\State \textbf{end for}
			\State \textbf{return} $\alpha^\star[n]$.
		\end{algorithmic}
	\end{algorithm}
	\fi
	\subsubsection{State}
	Time frame-dependent variables, such as $i$, $l_i$, and $L_i$, are now re-expressed as time-dependent variables. For instance, $i[n]$ denotes the current time frame index (i.e., current aperture index): 
	\[
	i[n]=i[n-1]+\mathds{1}\{a[n-1]=1,\,a[n-2]=0\}, \forall n\geq 3, 
	\]
	with $ i[1]=i[2]=1$. Similarly, $l[n]$ denotes the most recent sensing time slot prior to current time slot $n$:
	\begin{equation}
		l[n]=l[n-1]+\left(n-1-l[n-1]\right)\mathds{1}\{a[n-1]=1\}, \forall n\geq 2,
	\end{equation}
	with $l[1]=1$. In the same spirit, $L[n]\delta_t$ denotes the accumulated \ac{sar} integration time: 
	\begin{equation}
		L[n]=	\left(L[n-1]+ 1\right)\mathds{1}\{a[n-1]=1\}, \forall n \geq 1,
	\end{equation}
	where $ L[1]=1$. Note that, for a horizon $N$, the frame-dependent optimization variables are recovered from their time-dependent counterparts as
	$I=i[N]$, $l_i=\max\limits_{n, \;i[n]=i}l[n]$, and $L_i=\max\limits_{n,\,i[n]=i}L[n],\;\forall i$.
	Accordingly, the system state in time slot $n$ is defined as 
	$\mathbf{s}[n]=(s_1[n],s_2[n],s_3[n],s_4[n],s_5[n])^\top$, where 
	$s_1[n]=L[n]$ denotes the current aperture length, 
	$s_2[n]=\mathbf{v}_e[l[n]]$ is the last estimated eavesdropper velocity, 
	$s_3[n]=r_e(n,l[n],L[n])$ is the current uncertainty radius, 
	$s_4[n]=\left|\theta_e[l[n]]-\theta_u[n]\right|$ is the angular separation between the estimated eavesdropper direction and the \ac{abs}, and 
	$s_5[n]=d_e[l[n]]-d_u[n]$ is the relative distance difference.
	\subsubsection{Reward} 
	The reward at time slot $n$ is designed to encourage high secrecy performance while enforcing sensing and communication constraints. During sensing slots (i.e., $a[n]=1$), no data transmission occurs and thus no secrecy gain is achieved. Accordingly, we set $r[n]=0$. For communication slots (i.e., $a[n]=0$), the reward is defined as\begin{equation}
		r[n]=
		\begin{dcases}
			-\rho_2, \quad  \quad \text{if} \quad\mathrm{SCR}(L[l[n]]) < \mathrm{SCR}_{\min},&\\
			R[n] - \rho_1\left[R_{\min}-\frac{1}{n}\sum_{i=1}^{n} R_u[i]\right]^+,  \text{otherwise},&
		\end{dcases}
	\end{equation}
	where $\rho_1$ and $\rho_2$ are positive penalty factors enforcing constraints (\ref{constraint:minimum_user_rate}) and (\ref{constraint:scr}), respectively. The first case penalizes the agent if communication is attempted before achieving sufficient sensing quality, while the second case balances secrecy-rate maximization and communication quality-of-service satisfaction.
	Note that constraints (\ref{constraint:time_frame}) and (\ref{constraint:aperture_length}) in problem ($\mathcal{P}$) are inherently satisfied through the \ac{mdp} design.		
	\subsection{Solution Algorithm}
	To solve problem $(\mathcal{P})$ based on the formulated \ac{mdp}, we adopt proximal policy optimization (PPO) \cite{ppo}. Let $\boldsymbol{\Phi}$ denote the policy parameters. The policy $\pi_{\boldsymbol{\Phi}}(a[n]|\textbf{s}[n])$ is optimized by maximizing the expected cumulative reward $J(\boldsymbol{\Phi})=\mathbb{E}_{\pi_{\boldsymbol{\Phi}}}\left[\sum_n \gamma^n r[n]\right]$, where $\gamma\in[0,1)$ is the discount factor.  The policy is updated via gradient ascent using the policy-gradient estimate $\nabla_{\boldsymbol{\Phi}} J(\boldsymbol{\Phi})=\mathbb{E}\left[\nabla_{\boldsymbol{\Phi}}\log \pi_{\boldsymbol{\Phi}}(a|\textbf{s})\hat{A}_n\right]$, where $\hat{A}_n$ denotes the advantage estimate \cite{ppo}. In each interaction step, the agent selects an action according to $\pi_{\boldsymbol{\Phi}}(a[n]|\textbf{s}[n])$. If communication is selected, the robust power-splitting factor $\alpha^\star[n]$ is computed using \textbf{Algorithm}~\ref{alg:alpha_search}. The resulting reward is then used to update the policy. We refer the reader to \cite{ppo} for implementation details of PPO.
	\section{Simulation Results}
	\begin{table}[]
		\centering
		\caption{System parameters \cite{robust_secure,amine1,amine2}.}
		\label{tab:my-table}
		\begin{adjustbox}{max width=\columnwidth}
			\begin{tabular}{|c|c?c|c?c|c|}
				\hline
				Parameter           & Value 					& Parameter & Value 						& Parameter &Value \\ \hline
				$N$ & $2.5\times 10^{3}$ 		&$\sigma_t$   &5 dBsm   						&$q_{u,y}$ & -20 m  \\ \hline	
				$M_t$ & 12    					&$\mathrm{SCR}_{\mathrm{min}}$ &10 dB  					&$\sigma_u^2$   & -50 dBm			 \\ \hline
				$\delta_t$ & 0.1                &$B_r$   &1 GHz	   								&$\sigma_e^2$	& -50 dBm    \\ \hline
				$r_r$ & 100 m   				&$\lambda_c$   & 0.12 m		        					&$v_{e,\mathrm{max}}$ &	 28 m/s \\ \hline
				$r_a$ & 200 m     					&$B_c$          &100 MHz 					&$a_{e,\mathrm{max}}$    & 2 m/s$^{2}$\\ \hline
				$v_a$ & 10 m/s       			&$\beta_0$  & -30 dB 	    					&$c$    & 3$\times10^{8}$ m/s	  \\ \hline
				$h$ & 100 m   					&$P_{\mathrm{com}}^{\mathrm{max}}$& 1 W						&$\epsilon$    & 1$\times10^{-2}$ \\ \hline
				$\lambda_r$ & 0.12 m 	 		&$R_{\mathrm{min}}$& 1 bits/s/Hz    				 &$\rho_1$    & 0.5 \\ \hline
				$\sigma_0$ & -5 dBsm 			&$q_{u,x}$ &  -50 m 							&$\rho_2$    & 0.5 \\ \hline
			\end{tabular}
		\end{adjustbox} 
	\end{table}

	This section presents simulation results for the proposed framework using the parameters in Table \ref{tab:my-table}. We provide open-source code with 	instructions, convergence plots, additional analysis, \ac{drl} hyper-parameters not listed in Table \ref{tab:my-table}, and trained models at https://gitlab.cs.fau.de/ok76owib/td-jsarc, which can be used to reproduce the present results. For generalization purposes, the agent is trained on different randomly generated eavesdropper trajectories as detailed in the repository. For comparison, we consider the following benchmark schemes:
	\begin{itemize}
		\item \textbf{Benchmark scheme 1 (Hybrid \ac{drl}):} 
		In this learning-based scheme, the agent jointly optimizes discrete time-allocation decisions and continuous power-splitting within a hybrid action space using RLlib \cite{rllib} \cite{rllib}. 
		\item \textbf{Benchmark scheme 2 (Equal-aperture time allocation):} 
		This is a non-learning scheme where all frames are assigned equal aperture lengths, i.e., $L_i=L,\;\forall i\in\mathcal{I}$. Parameters $L$ and $I$ are optimized via grid search to maximize performance under the given system constraints.
		\item \textbf{Benchmark scheme 3 (Random time allocation):} 
		In this scheme, aperture lengths are randomly generated while satisfying constraint (\ref{constraint:scr}). 
		The power-splitting factor $\alpha[n]$ is computed using \textbf{Algorithm}~\ref{alg:alpha_search}. 
		The performance is evaluated by averaging over $10^3$ realizations.
	\end{itemize}
	\begin{figure}[t]
		\centering	
		\begin{subfigure}{\linewidth}
			\centering
			\ifonecolumn
			\includegraphics[width=5in]{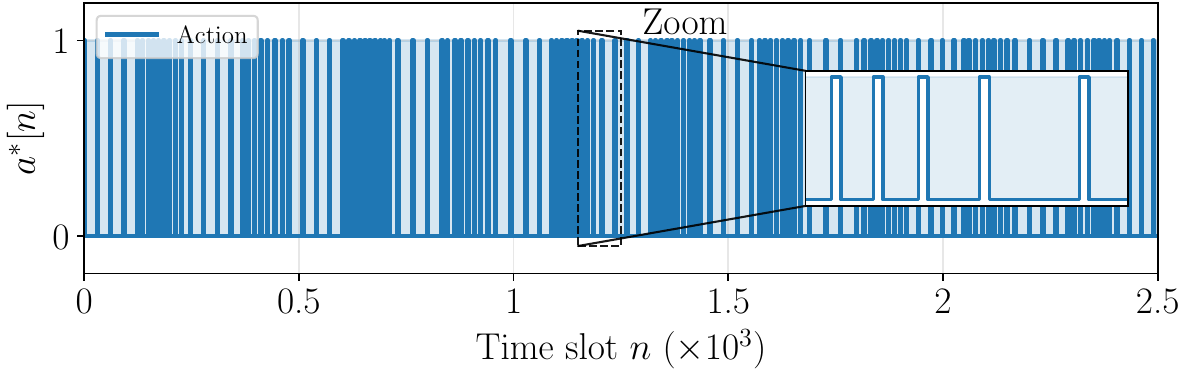}
			\fi
		\end{subfigure}
		\begin{subfigure}{\linewidth}
			\centering
			\ifonecolumn
			\includegraphics[width=5in]{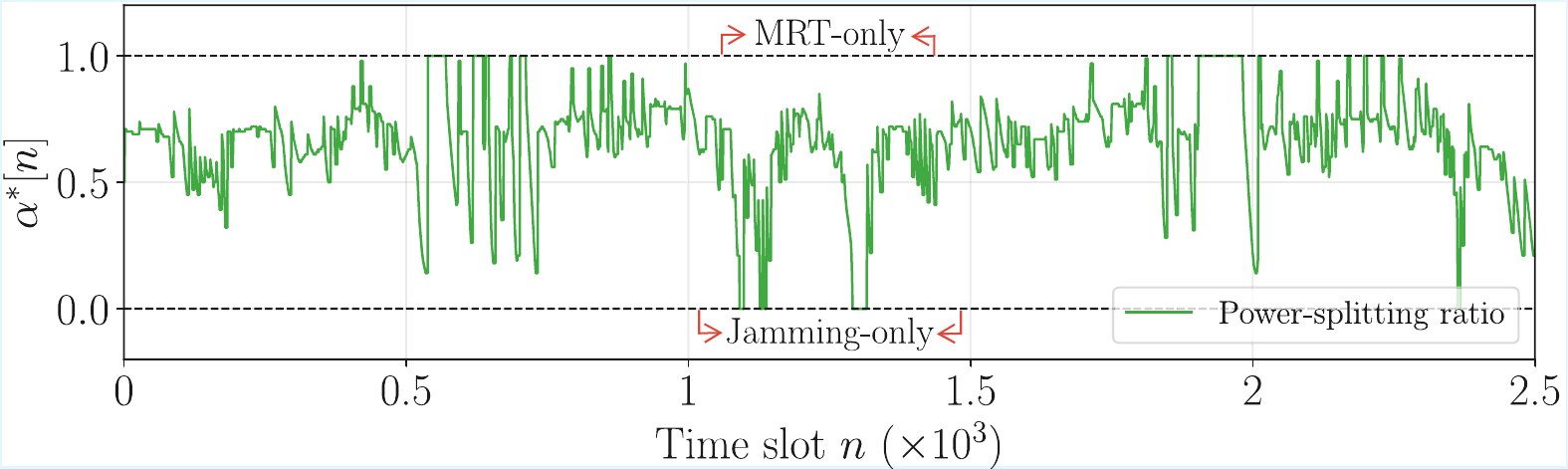}
			\fi
		\end{subfigure}
		\begin{subfigure}{\linewidth}
			\centering
			\ifonecolumn
			\includegraphics[width=5in]{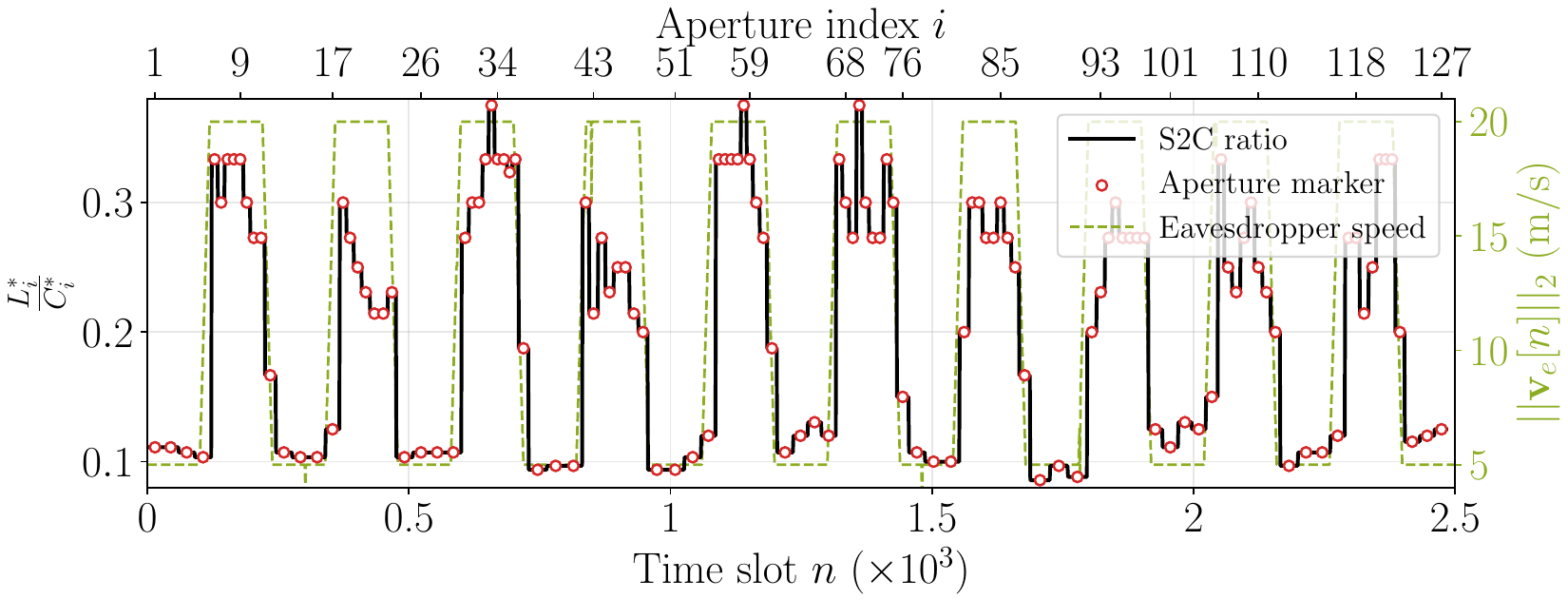}
			\fi
		\end{subfigure}
		\caption{Learned policy adaptation to unseen linear eavesdropper trajectory with oscillating speed (between $5$ and 	$20$~m/s). The top panel shows the binary action $a^{\star}[n]$. The middle panel depicts the optimal power-splitting factor $\alpha^{\star}[n]$, and the bottom panel shows the \ac{s2c} ratio across sensing apertures and the corresponding eavesdropper speed magnitude.}
		\label{fig:policy}
	\end{figure}

	In Fig.~\ref{fig:policy}, we investigate the impact of the eavesdropper’s speed on the learned policy. The eavesdropper follows a linear trajectory with oscillating speed between $5$ and $20$~m/s (see bottom panel), not considered during the actor–critic training. Nevertheless, the learned policy satisfies all constraints of problem $(\mathcal{P})$ and achieves an average worst-case secrecy rate of $5.25$~bits/s/Hz. From the top panel, we observe that sensing is activated more frequently during high-speed intervals, while communication dominates during slower phases. This reflects the need for more frequent sensing updates when the eavesdropper moves faster. The bottom panel further illustrates this via the \ac{s2c} ratio, which closely follows the oscillatory speed profile. This is consistent with \eqref{eq:uncertainty_radius}, as faster motion leads to a larger uncertainty radius $r_e$. The middle panel shows the power-splitting factor computed via \textbf{Algorithm}~\ref{alg:alpha_search}. The values adapt dynamically to the instantaneous geometry, balancing \ac{mrt} beamforming and jamming based on the relative distances and azimuth angles between the ABS, user, and eavesdropper.\par
\begin{figure}
		\centering
		\ifonecolumn
		\includegraphics[width=4in]{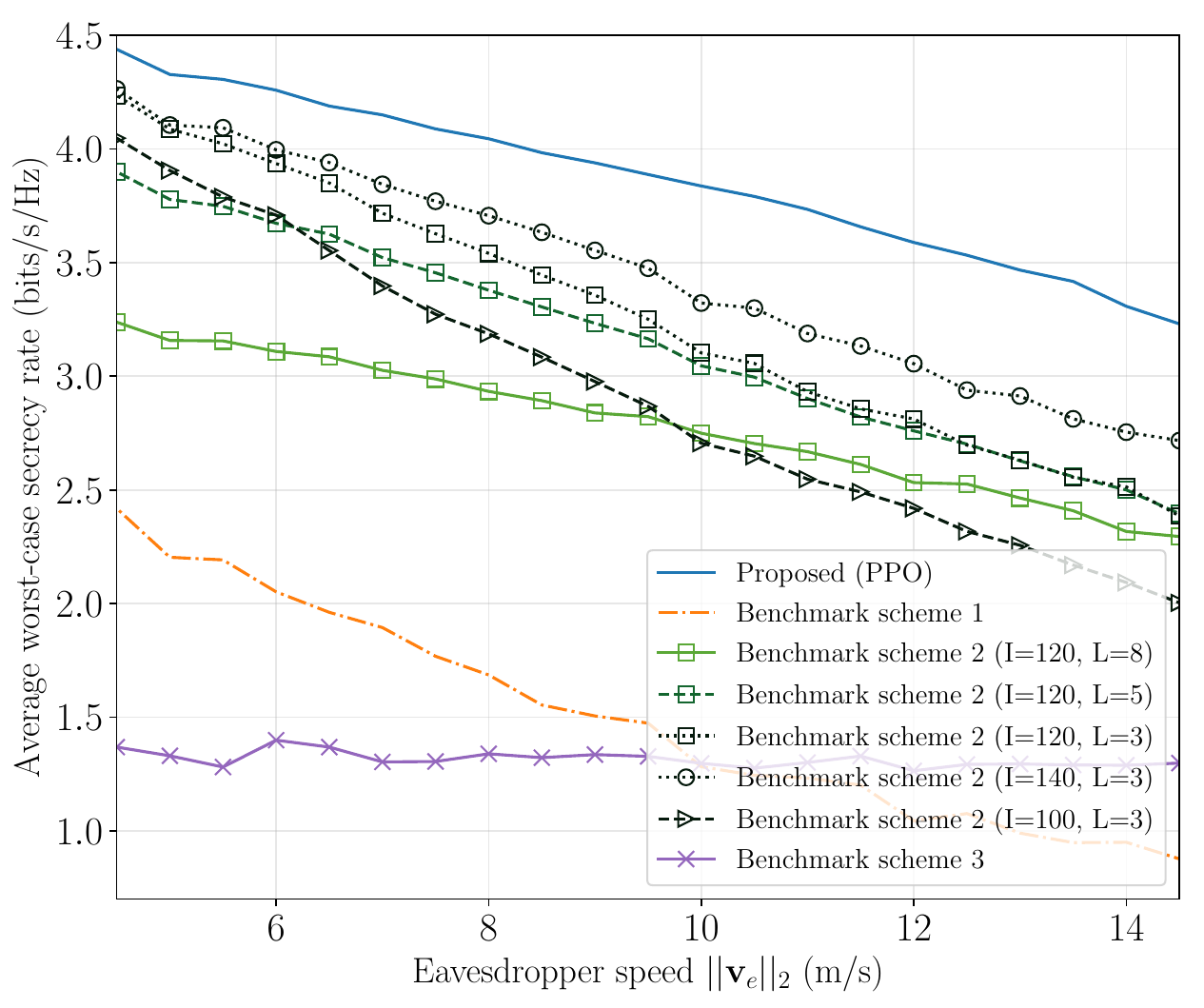}
		\fi
		\caption{Average worst-case secrecy rate versus eavesdropper speed. The eavesdropper moves along a circular trajectory centered at the user with radius $55$~m.}
		\label{fig:rate_vs_speed}
	\end{figure}

 
	In Fig.~\ref{fig:rate_vs_speed}, we compare the proposed framework with the benchmark schemes for different eavesdropper speeds. The eavesdropper moves along a circular trajectory centered at the user with a radius of $55$~m and constant speed. The proposed method consistently achieves the highest average worst-case secrecy rate across the entire speed range. The performance gap becomes more pronounced as the eavesdropper speed increases. For example, at $\|\mathbf{v}_e\|_2=14$~m/s, the proposed scheme achieves approximately $3.3$~bits/s/Hz compared to about $2.7$~bits/s/Hz for the best benchmark. Hybrid DRL underperforms due to the joint discrete–continuous learning complexity, which was simplified in the proposed solution using \textbf{Algorithm}~\ref{alg:alpha_search}.
	Benchmark scheme 2 is moderately effective but sensitive to the fixed values of $I$ and $L$, highlighting the limits of static time allocation for unknown eavesdropper trajectories, whereas Benchmark~3 performs worst due to the random time allocation.
	\section{Conclusion}
	In this paper, we proposed a \ac{td} \ac{jsarc} scheme that exploits cognitive \ac{sar} \ac{ati} to first estimate the location of a ground-moving eavesdropper and subsequently enables secure downlink communication with a \ac{ue} through adaptive beamforming and jamming. The resulting joint time and power allocation problem is addressed using a learning-based approach that combines discrete scheduling with robust power allocation optimization.  Simulation results demonstrated that the proposed solution significantly improves the secrecy rate compared to learning- and non-learning-based benchmark schemes, while maintaining robustness to previously unseen eavesdropper trajectories. Future work will extend the proposed framework to joint trajectory, beamforming, and jamming optimization under more general \ac{sar}-\ac{ati} constraints and explore higher \ac{sar}-communication integration, e.g., via waveform design.
		\vspace{-1mm}
	\bibliographystyle{IEEEtran}
	\bibliography{biblio}
	
\end{document}